\documentclass[journal]{IEEEtran}
\ifCLASSINFOpdf
\else
\fi
\ifCLASSOPTIONcompsoc
 \usepackage[caption=false,font=normalsize,labelfont=sf,textfont=sf]{subfig}
\else
 \usepackage[caption=false,font=footnotesize]{subfig}
\fi
\begin{document}
%
\title{GritNet 2: Real-Time Student Performance Prediction with Domain Adaptation}
%
%
%

\author{Byung-Hak Kim,
        Ethan Vizitei,
        and~Varun Ganapathi
\thanks{BH Kim, E Vizitei, V Ganapathi are with Udacity, Mountain View,
CA, 94040 USA (e-mail: \{hak, ethan, varun\}@udacity.com).}
\thanks{}}

%
%

\markboth{}%
{Shell \MakeLowercase{\textit{et al.}}: Bare Demo of IEEEtran.cls for IEEE Journals}
%



\maketitle

\begin{abstract}
Increasingly fast development and update cycle of online course contents, and diverse demographics of students in each online classroom, make student performance prediction in real-time (before the course finishes) and/or on curriculum without specific historical performance data available interesting topics for both industrial research and practical needs. In this research, we tackle the problem of real-time student performance prediction with on-going courses in a domain adaptation framework, which is a system trained on students' labeled outcome from one set of previous coursework but is meant to be deployed on another. In particular, we first introduce recently-developed GritNet architecture which is the current state of the art for student performance prediction problem, and develop a new \emph{unsupervised} domain adaptation method to transfer a GritNet trained on a past course to a new course without any (students' outcome) label. Our results for real Udacity students' graduation predictions show that the GritNet not only \emph{generalizes} well from one course to another across different Nanodegree programs, but enhances real-time predictions explicitly in the first few weeks when accurate predictions are most challenging. 
\end{abstract}


%
\IEEEpeerreviewmaketitle

\section{Introduction}
\label{sec:introduction}
%
%
%
%
With the growing need for people to keep learning throughout their careers, massive open online course (MOOCs) companies, such as Udacity and Coursera, not only aggressively design new courses that are relevant 
but refresh existing courses' content frequently to keep them up-to-date. This effort results in a significant increase in student numbers, which makes it impractical for even experienced human instructors to assess an individual student
and anticipate their learning outcomes. Moreover, students in each MOOC classroom are heterogeneous in their background and intention, which is very different from a classic classroom \cite{Chuang16, Economist17}. Even subsequent offerings of a course within a year can have a different population of students, mentors, and - in some cases - instructors. Also, due to the nascent nature of the online learning platforms, many other aspects of a course evolve quickly so that students are frequently being exposed to experimental content modalities or workflow refinements. In this world of MOOCs, an automated machine which reliably forecasts students' performance early in their coursework would be a valuable tool for making smart decisions about when (and with whom) to make live educational interventions, with the aim of increasing engagement, providing motivation, and empowering students to succeed. 

In the last three years, a few deep learning based neural network models for predicting students' future performance have been explored within data mining and learning analytic communities \cite{Mi15, Piech15, Whitehill17, Wang17, Kim18}. Since the majority of prior research on student performance prediction has focused on when both training (source) and testing (target) course data is sampled from the same course, 
the challenge of predicting student performance on recently created curriculum (for which there is no historical student performance data available) and/or the problems of real-time prediction during a live or on-going course (not after it has finished) have been studied in very little literature. While \cite{Boyer15} reports preliminary transfer learning performance, they also emphasize that MOOCs would be an ideal (and paramount) use case for transfer learning, in learning from the previous course to make predictions and design interventions in the next course.  In \cite{Whitehill17, Whitehill17b}, the authors assert that accuracy estimates obtained by training on the same course as the target course for deployment (generally not possible in real-world intervention scenarios) could be overly optimistic. Both works explore an idea of \emph{proxy labels} (in-situ) as an approximate quantity of interest which can be collected before a course completes by looking at whether each student interacted with the coursework at least once in the last week. 

In this paper, we first recast the student performance prediction problem as a sequential event prediction problem and introduce the current state of the art architecture GritNet in Section~\ref{sec:GritNet}. Then Section~\ref{sec:Domain Adaptation with GritNet} is devoted to developing a novel \emph{unsupervised} domain adaptation method with the GritNet. Specifically, in order to leverage the unlabeled data in the target course, we first propose a (simple) ordinal input encoding procedure as a basis for GritNet to provide transferable sequence-level embeddings across courses and use a trained GritNet on a source course to assign pseudo outcome labels to the target course, then continue training the GritNet on the target course. Unlike (crude) \emph{proxy labels} in \cite{Boyer15, Whitehill17, Whitehill17b}, these pseudo target labels are produced principally by using the source GritNet and also used to fine-tune the last fully-connected layer of the GritNet to be applied to other different courses. As far as we know, this is the first satisfying work to effectively answer the source and target course distribution mismatch encountered in real-time student performance prediction, which had been considered as a challenging open problem in prior works \cite{Boyer15, Dalipi18}. Finally, we provide, in Section~\ref{sec:Generalization Performance}, generalization performance results across (real) Udacity Nanodegree programs' data, and conclude in Section~\ref{sec:Conclusion}.

\section{GritNet}
\label{sec:GritNet}
The task of predicting student performance can be expressed as a sequential event prediction problem \cite{Rudin11}: given a past event sequence $\mathbf{o}\triangleq(o_{1},\dots,o_{T})$ taken by a student, estimate likelihood of future event sequence $\mathbf{y}\triangleq(y_{T+D},\dots,y_{T'})$ where $D \in \mathbb Z_{+}$. In the form of online classes, each event $o_{t}$ represents a student's action (or activities) associated with a time stamp. In other words, $o_{t}$ is defined as a paired tuple of $(a_{t},d_{t})$. Each action $a_{t}$ represents, for example, ``a lecture video viewed'', ``a quiz answered correctly/incorrectly'', or ``a project submitted and passed/failed'', and $d_{t}$ states the corresponding (logged) time stamp. Then, log-likelihood of $p(\mathbf{y}|\mathbf{o})$ can be written approximately as Equation~\ref{eq1}, given fixed-dimensional embedding representation $\upsilon$ of $\mathbf{o}$.
\begin{equation} \label{eq1}
\begin{split}
\log p(\mathbf{y}|\mathbf{o}) \simeq \sum_{i=T+D}^{T'} \log p(y_{i}|\upsilon)
\end{split}
\end{equation}
The goal of each GritNet is, therefore, to compute an individual log-likelihood $\log p(y_{i}|\upsilon)$, and those estimated scores can be simply added up to estimate long-term student outcomes.

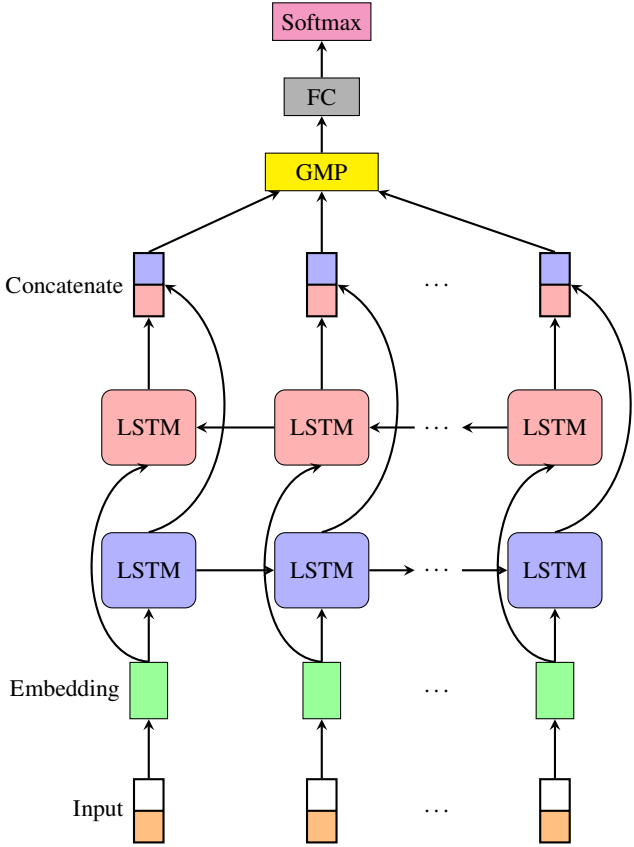
\begin{figure}[!t]
\vskip 0.2in
\usetikzlibrary{shapes.geometric, shapes.multipart, arrows, calc, positioning}

\tikzstyle{bw} = [rectangle, rounded corners, minimum width=1.25cm, minimum height=1cm,text centered, draw=black, fill=red!30]
\tikzstyle{fw} = [rectangle, rounded corners, minimum width=1.25cm, minimum height=1cm,text centered, draw=black, fill=blue!30]
\tikzstyle{emb} = [rectangle, minimum width=0.50cm, minimum height=0.75cm, text centered, draw=black, fill=green!40]
\tikzstyle{feat}=[rectangle split, rectangle split parts=2, rectangle split part fill={white!50, orange!50}, draw=black, thick]
\tikzstyle{cat}=[rectangle split, rectangle split parts=2, rectangle split part fill={blue!30, red!30}, draw=black, thick] 
\tikzstyle{agg} = [rectangle, minimum width=1.5cm, minimum height=0.5cm,text centered, draw=black, fill=yellow!100]
\tikzstyle{fc} = [rectangle, minimum width=1.0cm, minimum height=0.5cm,text centered, draw=black, fill=black!30]
\tikzstyle{softmax} = [rectangle, minimum width=1.0cm, minimum height=0.5cm,text centered, draw=black, fill=magenta!50]

\tikzstyle{arrow} = [thick,->,>=stealth]

\begin{tikzpicture}[node distance=1.5cm]

    \tikzstyle{every node}=[font=\small]

    \node (cat0) [cat, label={left:\textsc{C}oncatenate}]{};
    \node (cat1) [cat, right of=cat0, xshift=0.8cm] {};
    \node (cat2) [cat, right of=cat1, xshift=1.6cm] {};
    \node (dot_cat) at ($(cat1)!.5!(cat2)$) {\ldots};
    
    \node (agg) [agg, above of=cat1] {GMP};
    \node (fc) [fc, above of=agg, yshift=-0.5cm] {FC};
    \node (softmax) [softmax, above of=fc, yshift=-0.5cm] {Softmax};
    
    \node (bw0) [bw, below of=cat0, yshift=-0.4cm] {LSTM};
    \node (bw1) [bw, right of=bw0, xshift=0.8cm] {LSTM};
    \node (bw2) [bw, right of=bw1, xshift=1.6cm] {LSTM};
    \node (dot_bw) at ($(bw1)!.5!(bw2)$) {\ldots};
    
    \node (fw0) [fw, below of=bw0, yshift=-0.4cm] {LSTM};
    \node (fw1) [fw, right of=fw0, xshift=0.8cm] {LSTM};
    \node (fw2) [fw, right of=fw1, xshift=1.6cm] {LSTM};
    \node (dot_fw) at ($(fw1)!.5!(fw2)$) {\ldots};
    
    \node (emb0) [emb, below of=fw0, yshift=-0.1cm, label={left:\textsc{E}mbedding}] {};
    \node (emb1) [emb, right of=emb0, xshift=0.8cm] {};
    \node (emb2) [emb, right of=emb1, xshift=1.6cm] {};
    \node (dot_emb) at ($(emb1)!.5!(emb2)$) {\ldots};

    \node (feat0) [feat, below of=emb0, yshift=-0.1cm, label={left:\textsc{I}nput}] {};
    \node (feat1) [feat, right of=feat0, xshift=0.8cm] {};
    \node (feat2) [feat, right of=feat1, xshift=1.6cm] {};
    \node (dot_feat) at ($(feat1)!.5!(feat2)$) {\ldots};
    
    \draw [arrow] (agg) -- (fc);
    \draw [arrow] (fc) -- (softmax);
    
    \draw [arrow] (cat0.north) -- (agg);
    \draw [arrow] (cat1.north) -- (agg);
    \draw [arrow] (cat2.north) -- (agg);
    
    \draw [arrow] (bw0) -- (cat0);
    \draw [arrow] (bw1) -- (cat1);
    \draw [arrow] (bw2) -- (cat2);
    
    \draw [arrow] (bw2) -- (dot_bw);
    \draw [arrow] (dot_bw) -- (bw1);
    \draw [arrow] (bw1) -- (bw0);
    
    \draw [arrow] (fw0) -- (fw1);
    \draw [arrow] (fw0.north) to [out=15,in=-30] (cat0.east);
    \draw [arrow] (fw1) -- (dot_fw);
    \draw [arrow] (fw1.north) to [out=15,in=-30] (cat1.east);
    \draw [arrow] (dot_fw) -- (fw2);
    \draw [arrow] (fw2.north) to [out=15,in=-30] (cat2.east);
    
    \draw [arrow] (emb0) -- (fw0);
    \draw [arrow] (emb0.north) to [out=175,in=195] (bw0.south);
    \draw [arrow] (emb1) -- (fw1);
    \draw [arrow] (emb1.north) to [out=175,in=195] (bw1.south);
    \draw [arrow] (emb2) -- (fw2);
    \draw [arrow] (emb2.north) to [out=175,in=195] (bw2.south);
    
    \draw [arrow] (feat0) -- (emb0);
    \draw [arrow] (feat1) -- (emb1);
    \draw [arrow] (feat2) -- (emb2);

\end{tikzpicture}
\caption{Architecture of a GritNet for the student performance prediction problem.}
\label{GritNetArch}
\vskip -0.2in
\end{figure}

In order to feed students' raw event records into the GritNet, it is necessary to encode the time-stamped logs (ordered sequentially) into a sequence of fixed-length input vectors by \emph{one-hot encoding}. A one-hot vector $\mathbbm{1}(a_{t}) \in \{0,1\}^L$, where $L$ is the number of unique actions and only one $j$-th element could take the value 1 as:
\begin{equation} \label{eq4}
\begin{split}
\mathbbm{1}(a_{t})_{j}\triangleq
\begin{cases}
    1              & \text{if } j= a_{t}\\
    0              & \text{otherwise}
\end{cases},
\end{split}
\end{equation}
is used to distinguish each activity $a_{t}$ from every other. Then, one-hot vectors of the same student are connected into a long vector sequence to represent the student's whole sequential activities in $\mathbf{o}$. To further capture students' varying learning speeds, GritNet defines the discretized time difference $\Delta_{t}$ between adjacent events as\footnote{For the Udacity data described in Section~\ref{ssec:Udacity Data}, we use day to represent inter-event time intervals.}: 
\begin{equation} \label{eq5}
\begin{split}
\Delta_{t} \triangleq d_{t}-d_{t-1}.
\end{split}
\end{equation}
Then, one-hot encode $\Delta_{t}$ into $\mathbbm{1}(\Delta_{t})$ and connect them with the corresponding $\mathbbm{1}(a_{t})$ to represent $\mathbbm{1}(o_{t})$ as:
\begin{equation} \label{eq6}
\begin{split}
\mathbbm{1}(o_{t}) \triangleq [\mathbbm{1}(a_{t});\mathbbm{1}(\Delta_{t})].
\end{split}
\end{equation}
Lastly, the output sequences shorter than the maximum event sequence length (of a given training set) are pre-padded with all $\mathbf{0}$ vectors.

The complete GritNet architecture is illustrated in Fig.~\ref{GritNetArch}. The first embedding layer \cite{Bengio01} learns an embedding matrix $\mathbf{E}^o \in \mathbb{R}^{E\times|O|}$, where $E$ and $|O|$ are the embedding dimension and the number of unique events (i.e., input vector $\mathbbm{1}(o_{t})$ size), to convert an input vector $\mathbbm{1}(o_{t})$ onto a low-dimensional embedding $\mathbf{\upsilon}_{t}$ defined as: 
\begin{equation} \label{eq7}
\begin{split}
\mathbf{\upsilon}_{t} \triangleq \mathbf{E}^o\mathbbm{1}(o_{t}).
\end{split}
\end{equation}
This (dense) event embedding $\mathbf{\upsilon}_{t}$ is then passed into the bidirectional long short term memory (BLSTM) \cite{Graves05} and the output vectors are formed by concatenating each forward and backward direction outputs. Next, a global max pooling (GMP) layer \cite{Collobert08} is added to form a fixed dimension vector (independent of the input event sequence length) by taking the maximum over time (over the input event sequence). This GMP layer is able to capture the most relevant signals over the sequence and is also able to deal with the imbalanced nature of data\footnote{The signals could be contained in a few events in the event sequence. As input sequences consist of hundreds of events, information could get lost if we only deal with the last hidden state of BLSTM, therefore we use globally max-pooled representation of the hidden states over the entire sequence instead.}. One can view this GMP layer as a hard self-attention layer or can consider the GMP layer output generates a sequence-level embedding of the whole input event sequence. The GMP layer output is, ultimately, fed into a fully-connected (FC) layer, and a softmax (i.e., sigmoid) layer subsequently, to calculate the log-likelihood $\log p(y_{i}|\upsilon)$.

In particular, GritNet is the first deep learning architecture which successfully advances the state of the art by demonstrating substantial prediction accuracy improvements (particularly pronounced in the first few weeks when predictions are extremely challenging). In contrast to other works, the GritNet does not need any feature engineering (it can learn from raw input) and it can operate on any (raw) student event data associated with a time stamp even when highly imbalanced (see \cite{Kim18} for more details).


\begin{figure*}[ht]
\centering
\subfloat[Two students' event sequences with content IDs]{\usetikzlibrary{shapes.geometric}

\def\Households{User 1122}
\def\Firms{User 2214}

\def\DF{D_{F,t}} \def \DB {D_{B,t}} \def\Dividends{Dividends}
\def\NL{\mathit{NL}_{t}} \def\NewLoans{New loans}
\def\WB{\mathit{WB}_{t}} \def\Wages{Wages}
\def\SA{C_{t}} \def\Consumption{Consumption}
\def\INT{\mathit{INT}_t} \def\Interests{Interests}
\def\RL{\mathit {RL}_{t}} \def\PaidBackLoans{Paid back loans}

\begin{tikzpicture}[every node/.style={font=\normalsize,
  minimum height=0.35cm,minimum width=0.5cm},]

\node [matrix, very thin,column sep=0.4cm,row sep=0.01cm] (matrix) at (0,0) {
  & \node(0,0) (\Households) {}; & & \node(0,0) (\Firms) {}; & \\
  
  & \node(0,0) (\Households 0) {}; & & \node(0,0) (\Firms 0) {}; & & & \\
  \node(0,0) (t0 left) {}; & & & & & & \node(0,0) (t0 right) {};\\
  
  & \node(0,0) (\Households 1) {}; & \node(0,0) (\Dividends 1) {};
    & \node(0,0) (\Firms 1) {}; & & & \\
  \node(0,0) (t1 left) {}; & & & & & & \node(0,0) (t1 right) {};\\
  
  & \node(0,0) (\Households 2) {}; & \node(0,0) (\Dividends 2) {};
    & \node(0,0) (\Firms 2) {}; & & & \\
  \node(0,0) (t2 left) {}; & & & & & & \node(0,0) (t2 right) {};\\
  
  & \node(0,0) (\Households 3) {}; & \node(0,0) (\Dividends 3) {};
    & \node(0,0) (\Firms 3) {}; & & & \\
  \node(0,0) (t3 left) {}; & & & & & & \node(0,0) (t3 right) {};\\

  & \node(0,0) (\Households 4) {}; & \node(0,0) (\Dividends 4) {};
    & \node(0,0) (\Firms 4) {}; & & & \\
  \node(0,0) (t4 left) {}; & & & & & & \node(0,0) (t4 right) {};\\
  
  & \node(0,0) (\Households 5) {}; & \node(0,0) (\Dividends 5) {};
    & \node(0,0) (\Firms 5) {}; & & & \\
  \node(0,0) (t5 left) {}; & & & & & & \node(0,0) (t5 right) {};\\
  
  & \node(0,0) (\Households 6) {}; & \node(0,0) (\Dividends 6) {};
    & \node(0,0) (\Firms 6) {}; & & & \\
  \node(0,0) (t6 left) {}; & & & & & & \node(0,0) (t6 right) {};\\

  & \node(0,0) (\Households 7) {}; & \node(0,0) (\Dividends 7) {};
    & \node(0,0) (\Firms 7) {}; & & & \\
  \node(0,0) (t7 left) {}; & & & & & & \node(0,0) (t7 right) {};\\
  
  & \node(0,0) (\Households 8) {}; & \node(0,0) (\Dividends 8) {};
    & \node(0,0) (\Firms 8) {}; & & & \\
  \node(0,0) (t8 left) {}; & & & & & & \node(0,0) (t8 right) {};\\

  & \node(0,0) (\Households 9) {}; & \node(0,0) (\Dividends 9) {};
    & \node(0,0) (\Firms 9) {}; & & & \\
  \node(0,0) (t9 left) {}; & & & & & & \node(0,0) (t9 right) {};\\

  & \node(0,0) (\Households 10) {}; & \node(0,0) (\Dividends 10) {};
    & \node(0,0) (\Firms 10) {}; & & & \\
  \node(0,0) (t10 left) {}; & & & & & & \node(0,0) (t10 right) {};\\

  & \node(0,0) (\Households 11) {}; & \node(0,0) (\Dividends 11) {};
    & \node(0,0) (\Firms 11) {}; & & & \\
  \node(0,0) (t11 left) {}; & & & & & & \node(0,0) (t11 right) {};\\

  & \node(0,0) (\Households 12) {}; & \node(0,0) (\Dividends 12) {};
    & \node(0,0) (\Firms 12) {}; & & & \\
  \node(0,0) (t12 left) {}; & & & & & & \node(0,0) (t12 right) {};\\

  & \node(0,0) (\Households 13) {}; & \node(0,0) (\Dividends 13) {};
    & \node(0,0) (\Firms 13) {}; & & & \\
  \node(0,0) (t13 left) {}; & & & & & & \node(0,0) (t13 right) {};\\

  & \node(0,0) (\Households 14) {}; & \node(0,0) (\Dividends 14) {};
    & \node(0,0) (\Firms 14) {}; & & & \\
  \node(0,0) (t14 left) {}; & & & & & & \node(0,0) (t14 right) {};\\

  & \node(0,0) (\Households 15) {}; & \node(0,0) (\Dividends 15) {};
    & \node(0,0) (\Firms 15) {}; & & & \\
  \node(0,0) (t15 left) {}; & & & & & & \node(0,0) (t15 right) {};\\

  & \node(0,0) (\Households 16) {}; & \node(0,0) (\Dividends 16) {};
    & \node(0,0) (\Firms 16) {}; & & & \\
  \node(0,0) (t16 left) {}; & & & & & & \node(0,0) (t16 right) {};\\

  & \node(0,0) (\Households 17) {}; & \node(0,0) (\Dividends 17) {};
    & \node(0,0) (\Firms 17) {}; & & & \\
  \node(0,0) (t17 left) {}; & & & & & & \node(0,0) (t17 right) {};\\

  & \node(0,0) (\Households 18) {}; & \node(0,0) (\Dividends 18) {};
    & \node(0,0) (\Firms 18) {}; & & & \\
  \node(0,0) (t18 left) {}; & & & & & & \node(0,0) (t18 right) {};\\

  & \node(0,0) (\Households 19) {}; & & \node(0,0) (\Firms 19) {}; & & & \\
  
  & \node(0,0) (\Households 20) {}; & & \node(0,0) (\Firms 20) {}; & \\
};

\fill 
	(\Households) node[draw,fill=white] {\Households}
	(\Firms) node[draw,fill=white] {\Firms};

\draw [dotted] 
  (t0 left) -- (t0 right) node[right] {}
  (t3 left) -- (t3 right) node[right] {}
  (t6 left) -- (t6 right) node[right] {}
  (t7 left) -- (t7 right) node[right] {}
  (t9 left) -- (t9 right) node[right] {}
  (t12 left) -- (t12 right) node[right] {}
  (t13 left) -- (t13 right) node[right] {}
  (t15 left) -- (t15 right) node[right] {}
  (t17 left) -- (t17 right) node[right] {}
  (t18 left) -- (t18 right) node[right] {};

\draw [dashed] 
  (\Households) -- (\Households 20)
  (\Firms) -- (\Firms 20);

\draw
  (\Households 1) node[draw,circle,fill=blue!20] {} node {\tiny 1432}
  (\Households 2) node[draw,circle,fill=blue!20] {} node {\tiny 6164}
  (\Households 3) node[draw,circle,fill=blue!20] {} node {\tiny 8543}
  (\Households 4) node[draw,circle,fill=blue!20] {} node {\tiny 4689}
  (\Households 5) node[draw,circle,fill=yellow!20] {} node {\tiny 5473}
  (\Households 6) node[draw,circle,fill=blue!20] {} node {\tiny 5123}
  (\Households 7) node[draw,circle,fill=yellow!20] {} node {\tiny 6548}
  (\Households 8) node[draw,circle,fill=blue!20] {} node {\tiny 1546}
  (\Households 9) node[draw,circle,fill=blue!20] {} node {\tiny 7894}
  (\Households 10) node[draw,circle,fill=blue!20] {} node {\tiny 1546}
  (\Households 11) node[draw,circle,fill=blue!20] {} node {\tiny 7789}
  (\Households 12) node[draw,circle,fill=yellow!20] {} node {\tiny 9745}
  (\Households 13) node[draw,circle,fill=red!20] {} node {\tiny 554}
  (\Households 14) node[draw,circle,fill=red!20] {} node {\tiny 554}
  (\Households 15) node[draw,circle,fill=green!20] {} node {\tiny 554}
  (\Households 16) node[draw,circle,fill=blue!20] {} node {\tiny 5435}
  (\Households 17) node[draw,circle,fill=blue!20] {} node {\tiny 6548}
  (\Households 18) node[draw,circle,fill=yellow!20] {} node {\tiny 7844};

\draw
  (\Firms 1) node[draw,circle,fill=blue!20] {} node {\tiny 1432}
  (\Firms 7) node[draw,circle,fill=blue!20] {} node {\tiny 5123}
  (\Firms 11) node[draw,circle,fill=blue!20] {} node {\tiny 1546}
  (\Firms 13) node[draw,circle,fill=blue!20] {} node {\tiny 7789}
  (\Firms 16) node[draw,circle,fill=red!20] {} node {\tiny 554}
  (\Firms 17) node[draw,circle,fill=red!20] {} node {\tiny 554}
  (\Firms 18) node[draw,circle,fill=red!20] {} node {\tiny 554};

\end{tikzpicture}%
\label{fig:encoding-a}}
\hfil
\subfloat[Two students' event sequences with ordinal IDs]{\usetikzlibrary{shapes.geometric}

\begin{tikzpicture}[every node/.style={font=\normalsize,
  minimum height=0.35cm,minimum width=0.5cm},]

\def\Households{User 1122}
\def\Firms{User 2214}

\def\DF{D_{F,t}} \def \DB {D_{B,t}} \def\Dividends{Dividends}
\def\NL{\mathit{NL}_{t}} \def\NewLoans{New loans}
\def\WB{\mathit{WB}_{t}} \def\Wages{Wages}
\def\SA{C_{t}} \def\Consumption{Consumption}
\def\INT{\mathit{INT}_t} \def\Interests{Interests}
\def\RL{\mathit {RL}_{t}} \def\PaidBackLoans{Paid back loans}

\node [matrix, very thin,column sep=0.4cm,row sep=0.01cm] (matrix) at (0,0) {
  & \node(0,0) (\Households) {}; & & \node(0,0) (\Firms) {}; & \\
  
  & \node(0,0) (\Households 0) {}; & & \node(0,0) (\Firms 0) {}; & & & \\
  \node(0,0) (t0 left) {}; & & & & & & \node(0,0) (t0 right) {};\\
  
  & \node(0,0) (\Households 1) {}; & \node(0,0) (\Dividends 1) {};
    & \node(0,0) (\Firms 1) {}; & & & \\
  \node(0,0) (t1 left) {}; & & & & & & \node(0,0) (t1 right) {};\\
  
  & \node(0,0) (\Households 2) {}; & \node(0,0) (\Dividends 2) {};
    & \node(0,0) (\Firms 2) {}; & & & \\
  \node(0,0) (t2 left) {}; & & & & & & \node(0,0) (t2 right) {};\\
  
  & \node(0,0) (\Households 3) {}; & \node(0,0) (\Dividends 3) {};
    & \node(0,0) (\Firms 3) {}; & & & \\
  \node(0,0) (t3 left) {}; & & & & & & \node(0,0) (t3 right) {};\\

  & \node(0,0) (\Households 4) {}; & \node(0,0) (\Dividends 4) {};
    & \node(0,0) (\Firms 4) {}; & & & \\
  \node(0,0) (t4 left) {}; & & & & & & \node(0,0) (t4 right) {};\\
  
  & \node(0,0) (\Households 5) {}; & \node(0,0) (\Dividends 5) {};
    & \node(0,0) (\Firms 5) {}; & & & \\
  \node(0,0) (t5 left) {}; & & & & & & \node(0,0) (t5 right) {};\\
  
  & \node(0,0) (\Households 6) {}; & \node(0,0) (\Dividends 6) {};
    & \node(0,0) (\Firms 6) {}; & & & \\
  \node(0,0) (t6 left) {}; & & & & & & \node(0,0) (t6 right) {};\\

  & \node(0,0) (\Households 7) {}; & \node(0,0) (\Dividends 7) {};
    & \node(0,0) (\Firms 7) {}; & & & \\
  \node(0,0) (t7 left) {}; & & & & & & \node(0,0) (t7 right) {};\\
  
  & \node(0,0) (\Households 8) {}; & \node(0,0) (\Dividends 8) {};
    & \node(0,0) (\Firms 8) {}; & & & \\
  \node(0,0) (t8 left) {}; & & & & & & \node(0,0) (t8 right) {};\\

  & \node(0,0) (\Households 9) {}; & \node(0,0) (\Dividends 9) {};
    & \node(0,0) (\Firms 9) {}; & & & \\
  \node(0,0) (t9 left) {}; & & & & & & \node(0,0) (t9 right) {};\\

  & \node(0,0) (\Households 10) {}; & \node(0,0) (\Dividends 10) {};
    & \node(0,0) (\Firms 10) {}; & & & \\
  \node(0,0) (t10 left) {}; & & & & & & \node(0,0) (t10 right) {};\\

  & \node(0,0) (\Households 11) {}; & \node(0,0) (\Dividends 11) {};
    & \node(0,0) (\Firms 11) {}; & & & \\
  \node(0,0) (t11 left) {}; & & & & & & \node(0,0) (t11 right) {};\\

  & \node(0,0) (\Households 12) {}; & \node(0,0) (\Dividends 12) {};
    & \node(0,0) (\Firms 12) {}; & & & \\
  \node(0,0) (t12 left) {}; & & & & & & \node(0,0) (t12 right) {};\\

  & \node(0,0) (\Households 13) {}; & \node(0,0) (\Dividends 13) {};
    & \node(0,0) (\Firms 13) {}; & & & \\
  \node(0,0) (t13 left) {}; & & & & & & \node(0,0) (t13 right) {};\\

  & \node(0,0) (\Households 14) {}; & \node(0,0) (\Dividends 14) {};
    & \node(0,0) (\Firms 14) {}; & & & \\
  \node(0,0) (t14 left) {}; & & & & & & \node(0,0) (t14 right) {};\\

  & \node(0,0) (\Households 15) {}; & \node(0,0) (\Dividends 15) {};
    & \node(0,0) (\Firms 15) {}; & & & \\
  \node(0,0) (t15 left) {}; & & & & & & \node(0,0) (t15 right) {};\\

  & \node(0,0) (\Households 16) {}; & \node(0,0) (\Dividends 16) {};
    & \node(0,0) (\Firms 16) {}; & & & \\
  \node(0,0) (t16 left) {}; & & & & & & \node(0,0) (t16 right) {};\\

  & \node(0,0) (\Households 17) {}; & \node(0,0) (\Dividends 17) {};
    & \node(0,0) (\Firms 17) {}; & & & \\
  \node(0,0) (t17 left) {}; & & & & & & \node(0,0) (t17 right) {};\\

  & \node(0,0) (\Households 18) {}; & \node(0,0) (\Dividends 18) {};
    & \node(0,0) (\Firms 18) {}; & & & \\
  \node(0,0) (t18 left) {}; & & & & & & \node(0,0) (t18 right) {};\\

  & \node(0,0) (\Households 19) {}; & & \node(0,0) (\Firms 19) {}; & & & \\
  
  & \node(0,0) (\Households 20) {}; & & \node(0,0) (\Firms 20) {}; & \\
};

\fill 
	(\Households) node[draw,fill=white] {\Households}
	(\Firms) node[draw,fill=white] {\Firms};

\draw [dotted] 
  (t0 left) -- (t0 right) node[right] {$Day\,0$}
  (t3 left) -- (t3 right) node[right] {$Day\,1$}
  (t6 left) -- (t6 right) node[right] {$Day\,2$}
  (t7 left) -- (t7 right) node[right] {$Day\,3$}
  (t9 left) -- (t9 right) node[right] {$Day\,4$}
  (t12 left) -- (t12 right) node[right] {$Day\,5$}
  (t13 left) -- (t13 right) node[right] {$Day\,6$}
  (t15 left) -- (t15 right) node[right] {$Day\,7$}
  (t17 left) -- (t17 right) node[right] {$Day\,8$}
  (t18 left) -- (t18 right) node[right] {$Day\,9$};

\draw [dashed] 
  (\Households) -- (\Households 20)
  (\Firms) -- (\Firms 20);

\draw
  (\Households 1) node[draw,circle,fill=blue!20] {} node {\tiny C:1}
  (\Households 2) node[draw,circle,fill=blue!20] {} node {\tiny C:2}
  (\Households 3) node[draw,circle,fill=blue!20] {} node {\tiny C:3}
  (\Households 4) node[draw,circle,fill=blue!20] {} node {\tiny C:4}
  (\Households 5) node[draw,circle,fill=yellow!20] {} node {\tiny Q:1}
  (\Households 6) node[draw,circle,fill=blue!20] {} node {\tiny C:5}
  (\Households 7) node[draw,circle,fill=yellow!20] {} node {\tiny Q:2}
  (\Households 8) node[draw,circle,fill=blue!20] {} node {\tiny C:6}
  (\Households 9) node[draw,circle,fill=blue!20] {} node {\tiny C:7}
  (\Households 10) node[draw,circle,fill=blue!20] {} node {\tiny C:6}
  (\Households 11) node[draw,circle,fill=blue!20] {} node {\tiny C:8}
  (\Households 12) node[draw,circle,fill=yellow!20] {} node {\tiny Q:3}
  (\Households 13) node[draw,circle,fill=red!20] {} node {\tiny P:1}
  (\Households 14) node[draw,circle,fill=red!20] {} node {\tiny P:1}
  (\Households 15) node[draw,circle,fill=green!20] {} node {\tiny P:1}
  (\Households 16) node[draw,circle,fill=blue!20] {} node {\tiny C:9}
  (\Households 17) node[draw,circle,fill=blue!20] {} node {\tiny C:10}
  (\Households 18) node[draw,circle,fill=yellow!20] {} node {\tiny Q:4};

\draw
  (\Firms 1) node[draw,circle,fill=blue!20] {} node {\tiny C:1}
  (\Firms 7) node[draw,circle,fill=blue!20] {} node {\tiny C:5}
  (\Firms 11) node[draw,circle,fill=blue!20] {} node {\tiny C:6}
  (\Firms 13) node[draw,circle,fill=blue!20] {} node {\tiny C:8}
  (\Firms 16) node[draw,circle,fill=red!20] {} node {\tiny P:1}
  (\Firms 17) node[draw,circle,fill=red!20] {} node {\tiny P:1}
  (\Firms 18) node[draw,circle,fill=red!20] {} node {\tiny P:1};

\end{tikzpicture}%
\label{fig:encoding-b}}
\caption{Illustration of two different students' event sequences enrolled in the same Udacity Nanodegree program - a student 1122 who eventually graduates and a dropout student 2214. \protect\subref{fig:encoding-a} Student 1122 first views four lecture videos (blue circles), then gets a quiz correct (yellow circle). In the subsequent 13 events, watches a series of videos, answers quizzes and submits the first project (ID 554) three times and finally gets it passed. Student 2214, who eventually drops out, views four lectures sparsely and fails the first project three times in a row. \protect\subref{fig:encoding-b} After grouping content IDs into three subcategories (content, quiz and project), by exploiting actual ordering implicit in the content paths within each category, each content ID is converted to corresponding ordinal ID.}
\label{fig:encoding}
\end{figure*}
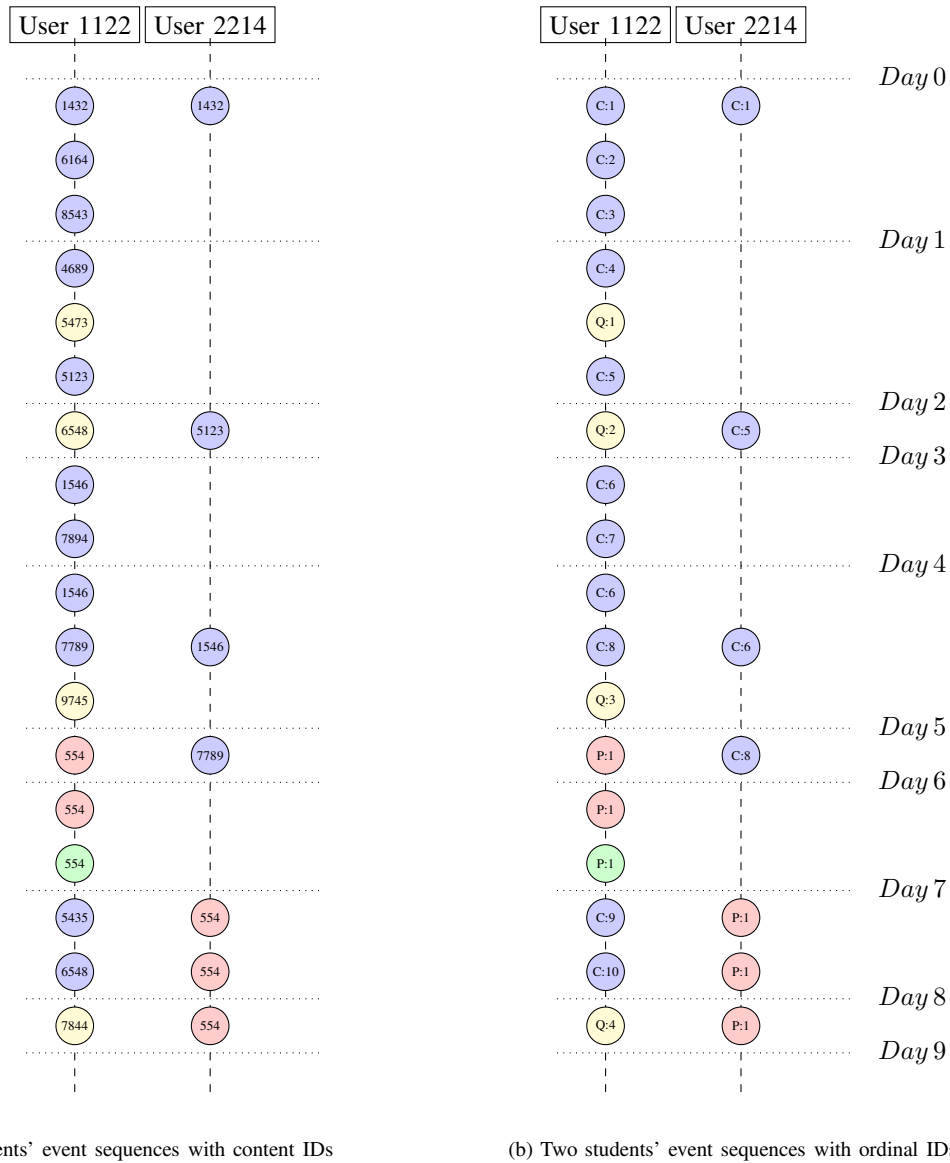

\section{Domain Adaptation with GritNet}
\label{sec:Domain Adaptation with GritNet}
Though GritNet's superior prediction accuracy was reported in \cite{Kim18}, it was not fully addressed whether GritNet models transfer well to different courses or if they could be deployed (or operationalized) for real-time or live predictions with on-going courses\footnote{The GritNet was trained and tested on the data from a specific Nanodegree program \emph{via} cross-validation. In that, the test data accuracy is a (overly optimistic) proxy for how it may perform on unseen courses.}. With an increase in new courses and the fast pace of content revisions of MOOC classes to meet students' educational needs, it requires the GritNet model trained on a previous course to generalize well to unseen courses (a new course or subsequent offering of the same course) and to predict performance in real-time, for students who have not yet finished the course. 
Performance in a real-time deployment scenario would be significantly impacted by such generalization. With that, in this section, we propose methods for overcoming the impact of the natural domain shift in the input data from a course over time into the GritNet model by leveraging domain adaptation techniques \cite{Pan10}.

\newcolumntype{C}{>{\raggedleft\arraybackslash}X} 
\begin{table*}
  \caption{Udacity Data Characteristics}
  \label{table:Udacity data characteristic}
  \centering
  \begin{tabularx}{\textwidth}{@{}l*{11}{C}c@{}}
    \toprule 
    Dataset & Enrolled From & Number of Students & Number of Contents (i) & Number of Quizzes (j) & Number of Projects (k) & Mean Seq. Length & Number of Graduates & Graduation Rates \\ 
    \midrule
    ND-A v1 & 2/3/2015 & 5,626   & 471 & 168 & 4 & 421 & 1,202 & 21.4\% \\ 
    ND-A v2 & 4/1/2015 & 2,230   & 471 & 168 & 4 & 881 & 453 & 20.3\% \\ 
    ND-B & 6/20/2016 & 13,639  & 514 & 287 & 10 & 285 & 2,180 & 16.0\% \\ 
    ND-C & 3/22/2017 & 4,377   & 568 & 84 & 10 & 675 & 1,726 & 39.4\% \\ 
    ND-D & 1/14/2017 & 4,761   & 346 & 50 & 5 & 430 & 2,198 & 46.2\%  \\ 
    \bottomrule
  \end{tabularx}
\end{table*}

\subsection{Ordinal Input Encoding}
\label{ssec:Ordinal Input Encoding}
Student activity records collected from different courses often have various lengths, formats and contents (for example, when the contents of a course are revised, new concepts are added or removed, or project requirements are changed for newly introduced courses), so that hand-designed aggregate features that are effective in one course might not be so in another\footnote{Even carefully engineered features' dimensions are usually constrained to be small and too restricted to tap the full benefits of sequential deep learning models and obviously had limited successful development of larger deep learning models in prior works \cite{Whitehill17,Mi15,Wang17}.}. To avoid those limitations, GritNet takes students' learning activities across time (e.g.,  Fig.~\ref{fig:encoding}~\subref{fig:encoding-a}) as a raw input sequence without feature engineering. 

For any MOOC coursework, each content exhibits a natural order, following a contents tree. For the sake of simplicity, we began our investigation by grouping content IDs into subcategories and normalizing them using natural ordering implicit in the content paths of the MOOC within each category\footnote{This simple encoding procedure would not require any modification of the GritNet architecture. We wish to generate a common mapping in which all the ordered but numeric (raw) contents IDs are converted into ordinal values for all online coursework. To ensure this, we could create a sophisticated encoding (e.g. table of contents), but this implies some additional complexity and maintenance of a common lookup source.}. This has performed well enough that we consider it a viable encoding for the moment. Then each encoded dataset contains the same number of unique actions as the original (e.g.,  Fig.~\ref{fig:encoding}~\subref{fig:encoding-b}). For example, Udacity data in Table~\ref{table:Udacity data characteristic} contains three subcategories - content pages, quizzes, and projects. Therefore, one Nanodegree program dataset has three sequences of ordinal IDs - content-$1$ to content-$i$, quiz-$1$ to quiz-$j$ and project-$1$ to project-$k$. Note that, with this encoding, the number of unique actions $L$ turns out to be equivalent to
\begin{equation} \label{eq8}
\begin{split}
L=i+2\times j+2\times k
\end{split}
\end{equation}
since quizzes and projects allow two potential actions (i.e., a quiz answered correctly/incorrectly or a project submitted and passed/failed). It is also worth mentioning that different courses may have different $i$, $j$, $k$ and $L$ values as shown in Table~\ref{table:Udacity data characteristic}.

\algnewcommand\algorithmicreturn{\textbf{return}}
\algnewcommand\RETURN{\algorithmicreturn}
\algnewcommand\algorithmicprocedure{\textbf{procedure}}
\algnewcommand\PROCEDURE{\item[\algorithmicprocedure]}%
\algnewcommand\algorithmicendprocedure{\textbf{end procedure}}
\algnewcommand\ENDPROCEDURE{\item[\algorithmicendprocedure]}%
\algnewcommand{\algvar}[1]{{\text{\ttfamily\detokenize{#1}}}}
\algnewcommand{\algarg}[1]{{\text{\ttfamily\itshape\detokenize{#1}}}}
\algnewcommand{\algproc}[1]{{\text{\ttfamily\detokenize{#1}}}}
\algnewcommand{\algassign}{\leftarrow}

\begin{algorithm}[ht]
    \caption{Domain Adaptation with GritNet}
    \label{alg:algorithm1}
    \begin{algorithmic}[1]
        \raggedright
        \REQUIRE Source course data $\mathcal{X}_{source}$, Source label $\mathcal{Y}_{source}$, Target data $\mathcal{X}_{target}$
        \STATE Set source training set as $\mathcal{T}_{source}=(\mathcal{X}_{source}, \mathcal{Y}_{source})$ \label{step1}
        \STATE Train ${\emph{GritNet}}_{source}$ with $\mathcal{T}_{source}$ \label{step2}
        \STATE Evaluate on $\mathcal{X}_{target}$: $\mathcal{\hat{Y}}_{pred}$:=${\emph{GritNet}}_{source}$($\mathcal{X}_{target}$) \label{step3}
        \STATE Assign pseudo-labels to $\mathcal{X}_{target}$: $\mathcal{Y}_{label}$:=${\mathbbm{1}}$($\mathcal{\hat{Y}}_{pred}\geq\theta$) \label{step4}
        \STATE Update target training set as $\mathcal{T}_{adapt}=(\mathcal{X}_{target}, \mathcal{Y}_{label})$ \label{step5}
        \STATE Freeze all the ${\emph{GritNet}}_{source}$ but the last FC layer \label{step6}
        \STATE Continue training with $\mathcal{T}_{adapt}$ \label{step7}
    \end{algorithmic}
\end{algorithm}

\subsection{Pseudo-Labels and Transfer Learning}
\label{ssec:Pseudo Labels and Transfer Learning}

Note that there are no target labels from the target course for training GritNet, and naive transfer from source GritNet to target course, in general, yields inferior performance (see results in Fig. \ref{fig:performance-1}). So to deal with unlabeled target course data in \emph{unsupervised} domain adaptation from source to target course, we harness the trained GritNet on a source data (which is a common real-world scenario when releasing new, never-before consumed educational content) to assign pseudo labels to the target data. 

Precisely, as described in Algorithm \ref{alg:algorithm1}, we use a trained GritNet to evaluate unannotated target data and assign hard one-hot labels if its prediction confidence exceeds a pre-determined threshold $\theta$ which is shown as Step~\ref{step4}. We use the softmax probability as a confidence measure. Given the pseudo-labels, we keep training a GritNet on the target course while freezing all the GritNet but the last FC layer as shown in Step~\ref{step6} and Step~\ref{step7}. This is to retrain only the top of the GritNet by continuing the backpropagation of errors on the target data which effectively fine-tunes the GritNet to the target course. 


This fine-tuning method was motivated by the observation that the GMP layer output captures generalizable abstract representation of input event sequence (i.e., sequence-level embedding), whereas the high-level FC layer learns the course-specific features. It is worth mentioning that the last FC layer limits the number of parameters to learn in Step~\ref{step7}. So this enables the entire Algorithm \ref{alg:algorithm1} to be a very practical solution for faster adaptation time (and, in fact, deployment cycle) and applicability for even a (relatively) smaller size target course\footnote{This approach is mainly to prevent overfitting even on smaller target datasets. Alternatively, we could fine-tune lower layers with smaller learning rate and increase it gradually as approaching to the FC layer which would lead to slower convergence. 
}.    

\section{Generalization Performance}
\label{sec:Generalization Performance}
\subsection{Udacity Data}
\label{ssec:Udacity Data}
See  Table~\ref{table:Udacity data characteristic} for detailed characteristics of each Udacity dataset used for this study. In all programs, graduation is defined as completing each of the required projects in a ND program curriculum with a passing grade. When a user officially graduates, their enrollment record is annotated with a time stamp, so it was possible to use the presence of this time stamp as the target label. Each program has a designated calendar start and end date for each cohort of students that passes through, and users have to graduate before the official end of their cohort's term to be considered successfully graduated. Each ND program's curriculum contains a mixture of video content, written content, quizzes, and projects. Note that it is not required to interact with every piece of content or complete every quiz to graduate. For all datasets, an event represents a user taking a specific action (e.g., watching a video, reading a text page, attempting a quiz, or receiving a grade on a project) at a certain time stamp. Some irrelevant data is filtered out during preprocessing, for example, events that occur \emph{before} a user's official enrollment as a result of a free-trial period. It should be noted that no personally identifiable information is included in this data and student equality is determined via opaque unique ids.

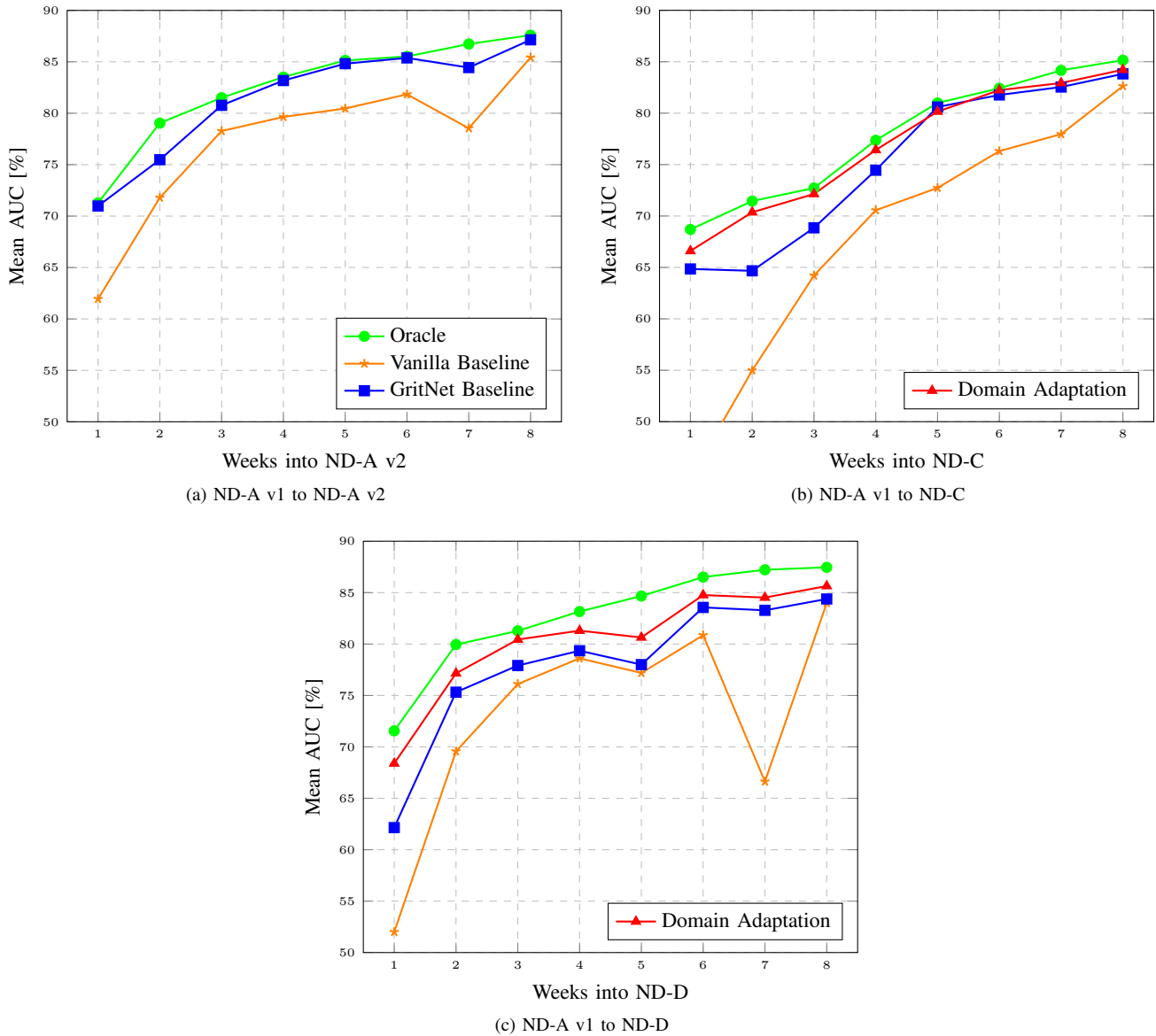
\begin{figure*}[!t]
\centering
\subfloat[ND-A v1 to ND-A v2]{\begin{tikzpicture}

\begin{axis}[
    xlabel={Weeks into ND-A v2}, 
    ylabel={Mean AUC [\%]},
    xmin=0.5, xmax=8.5, xtick={1,2,...,7,8},
    ymin=50, ymax=90, ytick={50,55,...,85,90},    
    xmajorgrids=true, 
    ymajorgrids=true, 
    grid style=dashed,
    label style={font=\small}, 
    tick label style={font=\tiny},  
    legend pos=south east,
    legend style={font=\small},
    legend cell align={left}
]

\addplot[color=green,mark=*,line width=0.8pt]
    coordinates {(1,71.28)
                (2,79.04)
                (3,81.51)
                (4,83.52)
                (5,85.13)
                (6,85.52)
                (7,86.73)
                (8,87.59)};

\addplot[color=orange,mark=star,line width=0.8pt]
    coordinates {(1,61.94)
                (2,71.79)
                (3,78.27)
                (4,79.64)
                (5,80.46)
                (6,81.84)
                (7,78.54)
                (8,85.42)};

\addplot[color=blue,mark=square*,line width=0.8pt]
    coordinates {(1,70.98)
                (2,75.48)
                (3,80.77)
                (4,83.18)
                (5,84.82)
                (6,85.38)
                (7,84.44)
                (8,87.14)};

\legend{Oracle, Vanilla Baseline, GritNet Baseline}
\end{axis}

\end{tikzpicture}%
\label{fig:performance-1a}}
\hfil
\subfloat[ND-A v1 to ND-C]{\begin{tikzpicture}

\begin{axis}[
    xlabel={Weeks into ND-C}, 
    ylabel={Mean AUC [\%]},
    xmin=0.5, xmax=8.5, xtick={1,2,...,7,8},
    ymin=50, ymax=90, ytick={50,55,...,95,100},
    xmajorgrids=true, 
    ymajorgrids=true, 
    grid style=dashed,
    label style={font=\small}, 
    tick label style={font=\tiny},  
    legend pos=south east,
    legend style={font=\small},
    legend cell align={left}
]

\addplot[color=green,mark=*,line width=0.8pt]
    coordinates {(1,68.69)
                (2,71.45)
                (3,72.72)
                (4,77.36)
                (5,80.99)
                (6,82.42)
                (7,84.16)
                (8,85.15)};

\addplot[color=orange,mark=star,line width=0.8pt]
    coordinates {(1,44.54)
                (2,54.98)
                (3,64.21)
                (4,70.56)
                (5,72.73)
                (6,76.31)
                (7,77.96)
                (8,82.63)};

\addplot[color=blue,mark=square*,line width=0.8pt]
    coordinates {(1,64.85)
                (2,64.67)
                (3,68.85)
                (4,74.45)
                (5,80.60)
                (6,81.77)
                (7,82.54)
                (8,83.82)};

\addplot[color=red,mark=triangle*,line width=0.8pt]
    coordinates {(1,66.60)
                (2,70.36)
                (3,72.14)
                (4,76.42)
                (5,80.15)
                (6,82.24)
                (7,82.93)
                (8,84.22)};

\legend{,,,Domain Adaptation}
\end{axis}

\end{tikzpicture}%
\label{fig:performance-1b}}
\hfil
\subfloat[ND-A v1 to ND-D]{\begin{tikzpicture}

\begin{axis}[
    xlabel={Weeks into ND-D}, 
    ylabel={Mean AUC [\%]},
    xmin=0.5, xmax=8.5, xtick={1,2,...,7,8},
    ymin=50, ymax=90, ytick={50,55,...,95,100},
    xmajorgrids=true, 
    ymajorgrids=true, 
    grid style=dashed,
    label style={font=\small}, 
    tick label style={font=\tiny},  
    legend pos=south east,
    legend style={font=\small},
    legend cell align={left}
]

\addplot[color=green,mark=*,line width=0.8pt]
    coordinates {(1,71.56)
                (2,79.95)
                (3,81.28)
                (4,83.16)
                (5,84.67)
                (6,86.51)
                (7,87.22)
                (8,87.46)};

\addplot[color=orange,mark=star,line width=0.8pt]
    coordinates {(1,52.00)
                (2,69.57)
                (3,76.11)
                (4,78.61)
                (5,77.19)
                (6,80.86)
                (7,66.63)
                (8,83.99)};

\addplot[color=blue,mark=square*,line width=0.8pt]
    coordinates {(1,62.15)
                (2,75.32)
                (3,77.91)
                (4,79.35)
                (5,78.00)
                (6,83.56)
                (7,83.28)
                (8,84.38)};

\addplot[color=red,mark=triangle*,line width=0.8pt]
    coordinates {(1,68.39)
                (2,77.15)
                (3,80.44)
                (4,81.30)
                (5,80.64)
                (6,84.76)
                (7,84.52)
                (8,85.64)};

\legend{,,,Domain Adaptation}

\end{axis}

\end{tikzpicture}%
\label{fig:performance-1c}}
\caption{Real-time student graduation prediction accuracy comparisons of models \protect\subref{fig:performance-1a} from ND-A v1 to the next version ND-A v2 and \protect\subref{fig:performance-1b}-\protect\subref{fig:performance-1c} from earlier ND-A v1 program to two later Udacity Nanodegree programs, ND-C and ND-D. \protect\subref{fig:performance-1a} GritNet baseline shows only 1.01\% abs accuracy loss (< 5.30\% abs loss of vanilla baseline) in average over eight weeks as compared to oracle bound. This accuracy gains of GritNet baseline suggests that sequence-level embedding trained with GritNet is more transferable features as compared to features learned with Vanilla baseline model. \protect\subref{fig:performance-1b}-\protect\subref{fig:performance-1c} Being consistent with \protect\subref{fig:performance-1a} results, clear wins of GritNet baseline over vanilla baseline (mostly pronounced at week 7 performance in \protect\subref{fig:performance-1c}) reaffirm that sequence-level embedding trained with GritNet is more robust to source and target course distribution mismatch as compared to features learned with Vanilla baseline model. Furthermore, domain adaptation provides 70.60\% accuracy recovery in average during first four weeks (up to 84.88\% at week 3) for ND-C dataset and 58.06\% accuracy recovery during the same four weeks (up to 75.07\% at week 3) for ND-D dataset from GritNet baseline performances.}
\label{fig:performance-1}
\end{figure*}

\begin{figure*}[!t]
\centering
\subfloat[ND-B to ND-C]{\begin{tikzpicture}

\begin{axis}[
    xlabel={Weeks into ND-C}, 
    ylabel={Mean AUC [\%]},
    xmin=0.5, xmax=8.5, xtick={1,2,...,7,8},
    ymin=50, ymax=90, ytick={50,55,...,95,100},
    xmajorgrids=true, 
    ymajorgrids=true, 
    grid style=dashed,
    label style={font=\small}, 
    tick label style={font=\tiny},  
    legend pos=south east,
    legend style={font=\small},
    legend cell align={left}
]

\addplot[color=green,mark=*,line width=0.8pt]
    coordinates {(1,68.39)
                (2,70.20)
                (3,72.74)
                (4,77.37)
                (5,80.82)
                (6,82.62)
                (7,83.92)
                (8,85.31)};

\addplot[color=blue,mark=square*,line width=0.8pt]
    coordinates {(1,54.17)
                (2,67.37)
                (3,69.65)
                (4,76.74)
                (5,80.46)
                (6,80.73)
                (7,83.00)
                (8,84.93)};

\addplot[color=red,mark=triangle*,line width=0.8pt]
    coordinates {(1,59.58)
                (2,70.71)
                (3,71.04)
                (4,77.60)
                (5,80.30)
                (6,80.76)
                (7,83.53)
                (8,85.25)};

\legend{Oracle, GritNet Baseline, Domain Adaptation}
\end{axis}

\end{tikzpicture}%
\label{fig:performance-2a}}
\hfil
\subfloat[ND-B to ND-D]{\begin{tikzpicture}

\begin{axis}[
    xlabel={Weeks into ND-D}, 
    ylabel={Mean AUC [\%]},
    xmin=0.5, xmax=8.5, xtick={1,2,...,7,8},
    ymin=50, ymax=90, ytick={50,55,...,95,100},    
    xmajorgrids=true, 
    ymajorgrids=true, 
    grid style=dashed,
    label style={font=\small}, 
    tick label style={font=\tiny},  
    legend pos=south east,
    legend style={font=\small},
    legend cell align={left}
]

\addplot[color=green,mark=*,line width=0.8pt]
    coordinates {(1,71.65)
                (2,79.65)
                (3,81.39)
                (4,82.33)
                (5,84.66)
                (6,86.71)
                (7,87.27)
                (8,87.48)
};

\addplot[color=blue,mark=square*,line width=0.8pt]
    coordinates {(1,59.44)
                (2,76.39)
                (3,78.21)
                (4,78.42)
                (5,82.26)
                (6,85.33)
                (7,85.35)
                (8,85.57)
};

\addplot[color=red,mark=triangle*,line width=0.8pt]
    coordinates {(1,62.90)
                (2,77.58)
                (3,78.28)
                (4,78.87)
                (5,83.12)
                (6,85.39)
                (7,85.73)
                (8,86.11)
};

\legend{Oracle, GritNet Baseline, Domain Adaptation}
\end{axis}

\end{tikzpicture}%
\label{fig:performance-2b}}
\caption{Real-time student graduation prediction accuracy comparisons of models from earlier ND-B to two later Udacity Nanodegree programs: \protect\subref{fig:performance-2a} ND-C and \protect\subref{fig:performance-2b} ND-D. Benefits of domain adaptation are less pronounced as compared to Fig.~\ref{fig:performance-1}~\protect\subref{fig:performance-1b} and Fig.~\ref{fig:performance-1}~\protect\subref{fig:performance-1c} results. This is because ND-B has much shorter average event sequence length than ND-A v1 (285 vs. 421) so that there is far less overlap with typical ND-C and ND-D event sequences. Even with those limitations in the source dataset ND-B, domain adaptation still yields 100.00\% accuracy recovery at week 2 for ND-C dataset and 36.41\% and 35.58\% accuracy recovery at week 2 and 5 respectively for ND-D dataset.}
\label{fig:performance-2}
\end{figure*}
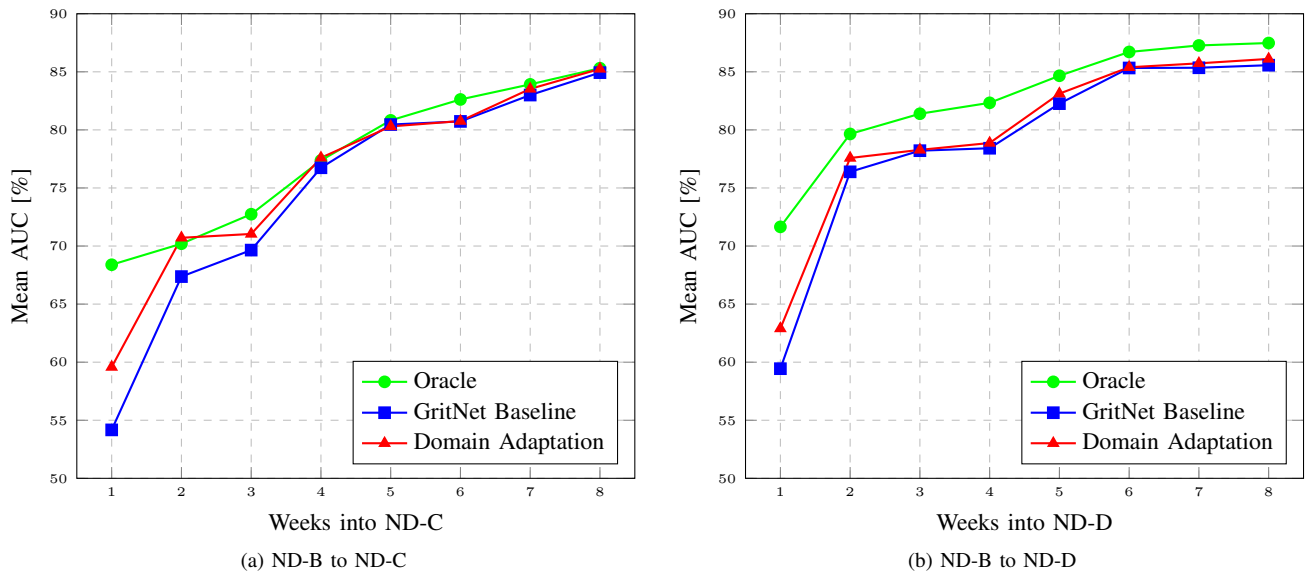

\subsection{Experimental Setup}
\label{ssec:Experimental Setup}
To demonstrate the benefits of Algorithm~\ref{alg:algorithm1}, we benchmark on the student graduation prediction task and compare various setups below:

\begin{itemize}
    \item \textbf{Vanilla Baseline}: In order to assess how much added value is brought by the GritNet, the same logistic regression based baseline model in \cite{Kim18} is trained on a source course and evaluated on a target one. 
    \item \textbf{GritNet Baseline}: We also train a GritNet based on a source course and evaluate performance on a target course. Note that there is no training data and labels from the target course for training the GritNet. 
    \item \textbf{Domain Adaptation}: We follow the steps specified in Algorithm~\ref{alg:algorithm1}. Note that for pseudo-labels generation (Step~\ref{step4}), a hyper-parameter $\theta$ threshold of $[0.1, 0.2, 0.3,0.4]$ is tried.  
    \item \textbf{Oracle}: To get a sense of the best possible performance limit of domain adaption, we carry out the same procedures as in Algorithm~\ref{alg:algorithm1} while skipping Step~\ref{step3} and Step~\ref{step4} to use oracle target labels $\mathcal{Y}_{target}$ instead of assigned pseudo-labels $\mathcal{Y}_{label}$.
\end{itemize}

\subsection{Training}
\label{Training}
We observed that the GritNet models are fairly easy to train. The training objective is the negative log likelihood of the observed event sequence of student activities under the model. The binary cross entropy loss is minimized\footnote{In this case, minimizing the binary cross entropy is equivalent to maximizing the log likelihood.} using RMSProp optimizer \cite{Hinton12} on a mini-batch size of 16. In our experiment, since the focus of the paper is to determine the benefits of domain adaption, we use the same set of hyperparameters across tasks and weeks. Specifically, single GritNet architecture of BLSTM with forward and backwards LSTM layers containing 256 cell dimensions per direction and embedding layer dimension of 512 are used. For all models across setups, we trained a different model for different weeks, based on students' week-by-week event records, to predict whether each student was likely to graduate. In that, further hyper-parameter optimization could be done for the optimal accuracy at each task and week. 

\subsection{Evaluation Measure}
\label{Evaluation Measure}
We used the Receiver Operating Characteristic (ROC) for evaluating the quality of predictions\footnote{Since the true binary target label (1: graduate, 0: not graduate) is imbalanced (i.e., number of 0s outweighs number of 1s), accuracy is not an appropriate metric.}. An ROC curve was created by plotting the true positive rate (TPR) against the false positive rate (FPR). In this task, the TPR is the percentage of students who graduate, which the GritNet labels positive, and the FPR is the percent of students who do not graduate, which the GritNet incorrectly labels positive. The accuracy of each system's prediction was measured by the area under the ROC curve (AUC) which scores between 0 and 100\% (the higher, the better) $-$ with random guess yielding 50\% all the time\footnote{Furthermore, AUC is commonly used in knowledge tracing and dropout prediction papers so we report all results as AUC for consistency.}. We used $5$-fold student level cross-validation, while ensuring each fold contained roughly the same proportions of the two groups (graduate and non graduate) of students.

To compare domain adaption performance with the upper bound assessed with oracle setup, we consider another performance measure, mean AUC recovery rate (ARR), as a metric to evaluate AUC improvements from \emph{unsupervised} domain adaptation against oracle training. This measure is defined as absolute AUC reduction that \emph{unsupervised} domain adaptation produces divided by the absolute reduction that oracle (\emph{supervised}) training produces as:

\begin{equation} \label{eq9}
\begin{split}
\text{ARR} \triangleq \frac{\text{UnsupAUC - BaselineAUC}}{\text{OracleAUC - BaselineAUC}}.
\end{split}
\end{equation}

We assume that the upper bound for \emph{unsupervised} domain adaptation would be the performance of the oracle training on the same data, which has a AUC recovery of 100\%.

\subsection{Performance Results}
\label{ssec:Performance Results}
In order to evaluate the transferability of the GritNet across different courses, we deliberately consider the following five representative real-world scenarios 
from a previous course to unseen courses (either subsequent offering of the same course or new courses)
\begin{itemize}
    \item \textbf{ND-A v1 to ND-A v2} (see Fig.~\ref{fig:performance-1}~\subref{fig:performance-1a}): There are minor contents changes between v1 and v2 of ND-A program. In this setting, accuracy gains of GritNet baseline (over vanilla baseline) as shown in Fig.~\ref{fig:performance-1}~\subref{fig:performance-1a} indicate that sequence-level embedding trained with GritNet is more transferable as compared to features learned with Vanilla baseline model. 
    \item \textbf{ND-A v1 to ND-C and ND-D} (see Fig.~\ref{fig:performance-1}~\subref{fig:performance-1b}-\subref{fig:performance-1c}): We analyze transferability of GritNet from an earlier ND-A v1 program to later ND-C and ND-D programs. ND-C curriculum has a lower expectation of prior technical knowledge than ND-D. 
    Being consistent with Fig.~\ref{fig:performance-1}~\subref{fig:performance-1a} results, clear wins of GritNet baseline over vanilla baseline in Fig.~\ref{fig:performance-1}~\subref{fig:performance-1b} and Fig.~\ref{fig:performance-1}~\subref{fig:performance-1c} results strongly confirm that sequence-level embedding trained with GritNet is more robust to source and target course distribution mismatch, as compared to features learned with the Vanilla baseline model. This signifies the sequence-level embedding is a sufficiently abstract input representation to generalize GritNet model across entirely different courses of study (especially to new ones where no student outcome data exists). In this respect, domain adaptation provides 70.60\% ARR in average during first four weeks (up to 84.88\% at week 3) for ND-C dataset and 58.06\% ARR during the same four weeks (up to 75.07\% at week 3) for ND-D dataset from GritNet baseline performances.
    Altogether, it suggests that all the benefits of domain adaptation in fact came from the retraining course-specific top FC layer while using the frozen lower layers for fixed sequence-level embedding.
    \item \textbf{ND-B to ND-C and ND-D} (see  Fig. \ref{fig:performance-2}): We consider another similar scenario from an earlier ND-B program to later ND-C and ND-D programs. Unlike previous ND-A v1, however, the ND-B dataset shows much shorter average event sequence length (285 vs. 421) so that there is far less overlap with typical target event sequences from ND-C and ND-D (285 vs. 675 and 430). In general, as a student progresses, the accumulated event sequence typically gets longer. So for those later weeks, there is less overlap, which in fact would limit inherent performance gains which could be achieved from domain adaptation. This is a primary reason why benefits of domain adaptation shown in Fig. \ref{fig:performance-2} are actually less observed as compared to Fig.~\ref{fig:performance-1} results. Even with those limitations in the source dataset ND-B, domain adaptation nevertheless yields 100.00\% ARR at week 2 for ND-C dataset and 36.41\% and 35.58\% ARR at week 2 and 5 respectively for ND-D dataset.
\end{itemize}

\section{Conclusion}
\label{sec:Conclusion}
In this paper, we have proposed an extremely practical approach for predicting real-time student performance, which has been considered as of the utmost importance in MOOCs (but under-explored), as this enables performance prediction while a course is on-going. Unlike prior works, we introduced a novel domain adaptation algorithm with GritNet and demonstrated that GritNet can be transferred to new courses without labels and provides a substantial AUC recovery rate. This method is effective in the sense that it works across different courses varying in lengths, format and contents and does not require custom feature engineering or additional target-course data or labels. Encouraged by this result, many future directions are feasible to explore. One potential direction is to look into a GritNet pretraining across more diverse courses. Given a pretrained GritNet and fine-tuning on the target course, it would boost performance further. We hope that our results will catalyze new developments of transfer learning for the real-time student performance prediction problem.

\ifCLASSOPTIONcaptionsoff
  \newpage
\fi



\bibliographystyle{IEEEtran}
%
\bibliography{tlt}

\begin{thebibliography}{10}
\providecommand{\url}[1]{#1}
\csname url@samestyle\endcsname
\providecommand{\newblock}{\relax}
\providecommand{\bibinfo}[2]{#2}
\providecommand{\BIBentrySTDinterwordspacing}{\spaceskip=0pt\relax}
\providecommand{\BIBentryALTinterwordstretchfactor}{4}
\providecommand{\BIBentryALTinterwordspacing}{\spaceskip=\fontdimen2\font plus
\BIBentryALTinterwordstretchfactor\fontdimen3\font minus
  \fontdimen4\font\relax}
\providecommand{\BIBforeignlanguage}[2]{{%
\expandafter\ifx\csname l@#1\endcsname\relax
\typeout{** WARNING: IEEEtran.bst: No hyphenation pattern has been}%
\typeout{** loaded for the language `#1'. Using the pattern for}%
\typeout{** the default language instead.}%
\else
\language=\csname l@#1\endcsname
\fi
#2}}
\providecommand{\BIBdecl}{\relax}
\BIBdecl

\bibitem{Chuang16}
I.~Chuang and A.~Ho, ``Harvard{X} and {MIT}x: Four years of open online
  courses, {F}all 2012-{S}ummer 2016,'' \emph{SSRN Electronic Journal}, 2016.

\bibitem{Economist17}
Economist, ``Equipping people to stay ahead of technological change -
  {L}ifelong learning,'' \emph{A Special Report On Lifelong Learning: How to
  {S}urvive in the {A}ge of {A}utomation}, 2017.

\bibitem{Mi15}
F.~Mi and D.-Y. Yeung, ``Temporal models for predicting student dropout in
  massive open online courses,'' in \emph{Proceedings of 15th IEEE
  International Conference on Data Mining Workshop (ICDMW 2015)}, Atlantic
  City, New Jersey, 2015, pp. 256--263.

\bibitem{Piech15}
C.~Piech, J.~Bassen, J.~Huang, S.~Ganguli, M.~Sahami, L.~Guibas, and
  J.~Sohl-Dickstein, ``Deep knowledge tracing,'' in \emph{Advances in Neural
  Information Processing Systems 28 (NIPS 2015)}, 2015, pp. 505--513.

\bibitem{Whitehill17}
J.~Whitehill, K.~Mohan, D.~Seaton, Y.~Rosen, and D.~Tingley, ``Delving deeper
  into {MOOC} student dropout prediction,'' \emph{arXiv preprint
  arXiv:1702.06404}, 2017.

\bibitem{Wang17}
W.~Wang, H.~Yu, and C.~Miao, ``Deep model for dropout prediction in {MOOCs},''
  in \emph{Proceedings of the 2nd International Conference on Crowd Science and
  Engineering (ICCSE 2017)}, Beijing, China, 2017, pp. 26--32.

\bibitem{Kim18}
B.-H. Kim, E.~Vizitei, and V.~Ganapathi, ``Grit{N}et: Student performance
  prediction with deep learning,'' in \emph{11th International Conference on
  Educational Data Mining (EDM 2018)}, 2018, pp. 625--629.

\bibitem{Boyer15}
S.~Boyer and K.~Veeramachaneni, ``Transfer learning for predictive models in
  massive open online courses,'' in \emph{16th International Conference on
  Artificial Intelligence in Education (AIED 2015)}.\hskip 1em plus 0.5em minus
  0.4em\relax Springer, 2015, pp. 54--63.

\bibitem{Whitehill17b}
J.~Whitehill, K.~Mohan, D.~Seaton, Y.~Rosen, and D.~Tingley, ``{MOOC} dropout
  prediction: How to measure accuracy?'' in \emph{Proceedings of the Fourth ACM
  Conference on Learning @ Scale (L@S 2017)}, Cambridge, Massachusetts, USA,
  2017, pp. 161--164.

\bibitem{Dalipi18}
F.~Dalipi, A.~S. Imran, and Z.~Kastrati, ``{MOOC} dropout prediction using
  machine learning techniques: Review and research challenges,'' in \emph{9th
  IEEE Global Engineering Education Conference (EDUCON 2018)}, 2018.

\bibitem{Rudin11}
C.~Rudin, B.~Letham, A.~Salleb-Aouissi, E.~Kogan, and D.~Madigan, ``Sequential
  event prediction with association rules,'' in \emph{24th Annual Conference on
  Learning Theory (COLT 2011)}, 2011, pp. 615--634.

\bibitem{Bengio01}
Y.~Bengio, R.~Ducharme, and P.~Vincent, ``A neural probabilistic language
  model,'' in \emph{Advances in Neural Information Processing Systems 13 (NIPS
  2000)}, 2001, pp. 932--938.

\bibitem{Graves05}
A.~Graves and J.~Schmidhuber, ``Framewise phoneme classification with
  bidirectional {LSTM} networks,'' in \emph{2005 International Joint Conference
  on Neural Networks (ICJNN'05)}, 2005, pp. 23--43.

\bibitem{Collobert08}
R.~Collobert and J.~Weston, ``A unified architecture for natural language
  processing: Deep neural networks with multitask learning,'' in \emph{25th
  International Conference on Machine Learning (ICML'08)}.\hskip 1em plus 0.5em
  minus 0.4em\relax ACM, 2008, pp. 160--167.

\bibitem{Pan10}
S.~J. Pan and Q.~Yang, ``A survey on transfer learning,'' \emph{IEEE
  Transactions on Knowledge and Data Engineering}, vol.~22, no.~10, pp.
  1345--1359, 2010.

\bibitem{Hinton12}
G.~Hinton, N.~Srivastava, and K.~Swersky, ``Lecture 6d - {A} separate, adaptive
  learning rate for each connection,'' \emph{{S}lides of {L}ecture {N}eural
  {N}etworks for {M}achine {L}earning}, 2012.

\end{thebibliography}
\end{document}